\documentclass[twocolumn,aps,prapplied]{revtex4-1}

\usepackage{amsfonts}
\usepackage{amsmath}
\usepackage{graphicx}
\usepackage{bm}
\usepackage{amssymb}
\usepackage{dcolumn}
\usepackage{color}
\usepackage{multirow}
\usepackage{booktabs}
\usepackage[colorlinks,
linkcolor=blue,
anchorcolor=blue,
citecolor=blue,
urlcolor=blue,
]{hyperref}
\begin{document}
\title{Great enhancement of Curie temperature and magnetic anisotropy in two-dimensional van der Waals magnetic semiconductor heterostructures}

\author{Xue-Juan Dong$^1$, Jing-Yang You$^1$, Zhen Zhang$^1$}

\author{Bo Gu$^{2,3}$}

\email[]{gubo@ucas.ac.cn}

\author{Gang Su$^{2,3,1}$}

\email[]{gsu@ucas.ac.cn}

\affiliation{$^1$ School of Physical Sciences, University of Chinese Academy of Sciences, Beijing 100049, China \\
$^2$ Kavli Institute for Theoretical Sciences,and CAS Center for Excellence in Topological Quantum Computation, University of Chinese Academy of Sciences, Beijing 100190, China \\
$^3$ Physical Science Laboratory, Huairou National Comprehensive Science Center, Beijing 101400, China}
     												
\date{\today}
\begin{abstract}
In two-dimensional (2D) magnetic systems, large magnetic anisotropy is needed  to stabilize the magnetic order according to Mermin-Wagner theorem. Based on density functional theory (DFT) calculations, we propose that the magnetic anisotropic energy (MAE) of 2D ferromagnetic (FM) semiconductors can be strongly enhanced in van der Waals heterostructures by attaching a nonmagnetic semiconductor monolayer with large spin-orbit coupling. We studied Cr$_2$Ge$_2$Te$_6$/PtSe$_2$ bilayer heterostructures, where each layer has been realized in recent experiments. The DFT calculations show that the MAE of Cr$_2$Ge$_2$Te$_6$/PtSe$_2$ is enhanced by 70\%, and the Curie temperature $T_C$ is increased far beyond room temperature. A model Hamiltonian is suggested to analyze the DFT results, showing that both the Dzyaloshinskii-Moriya interaction and the single-ion anisotropy contribute to the enhancement of the MAE. Based on the superexchange picture, we find that the decreased energy difference between 3d orbitals of Cr and 5p orbitals of Te contributes partially to the increase of $T_C$.  Our present work indicates a promising way to enhance the MAE and $T_C$ by constructing van der Waals semiconductor heterostructures, which will inspire further studies on the 2D magnetic semiconductor systems.
\end{abstract}
\maketitle

\section{Introduction}
To achieve room temperature ferromagnetic semiconductors has long been an important topic in science \cite{Kennedy2005}. Recently, magnetism in two-dimensional (2D) van der Waals materials has become a hot-spot which could lead to new device applications in information storage and spintronics \cite{nature-magreview2018,Miller2017,Sethulakshmi2019}. People have devoted to explore 2D van der Waals ferromagnetic semiconductors, such as Cr$_2$Ge$_2$Te$_6$ \cite{CGT-nature2017}, CrI$_3$ \cite{CrI3-2017}, Fe$_3$GeTe$_2$ \cite{zhangyuanbo2018}, VSe$_2$ \cite{Bonilla2018VSe2}, MnSe$_2$ \cite{OHara2018MnSe2}, and FePS$_3$ \cite{2019jpclFePS3,Lee2016FePS3}, etc. These layered materials have enriched the database of 2D ferromagnetic materials, but there are still challenges ahead. On one hand, large magnetic anisotropic energy (MAE) is required to suppress the thermal fluctuations and to stabilize the 2D magnetic moment according to Mermin-Wagner theorem \cite{Mermin1966}. Different pictures, for instance, Kitaev interaction, single-ion anisotropy (SIA) \cite{npjkitaev2018} and p-d covalency \cite{npjkitaev2018} have been suggested to understand the MAE in 2D van der Waals ferromagnetic semiconductors.  The Kitaev interaction and the SIA are originated from the spin-orbit coupling (SOC),  the latter as a relativistic effect is usually small in many materials. On the other hand, high Curie temperature $T_C$ above room temperature is very required from practical applications \cite{Li2019semi}. The Curie temperature of Cr$_2$Ge$_2$Te$_6$ bilayer and CrI$_3$ monolayer is 28 K and 45 K, respectively, far below room temperature \cite{CGT-nature2017,CrI3-2017}. Efforts have been paid to enhance the magnetization of 2D van der Waals magnets and to seek for new 2D magnetic material as well as to control the magnetism using mechanic and electronic means \cite{Jiang2018elec,You2019}. Strain is another useful way to control the magnetism in 2D ferromagnetic semiconductors. For example, $T_C$ in Cr$_2$Ge$_2$Se$_6$ monolayer can be increased beyond room temperature by applying just a few percent tensile strain \cite{Dong2019}, and the MAE of CrI$_3$ monolayer can also be tuned by strain \cite{2018prbstrainCrX3}.

It is known that combining two kinds of 2D materials into heterostructures could regulate the overall performance \cite{2019nature-vdw,2019nanotech,2018nanotech-ultrafast,Novoselov2016science}, which can thus provide platforms to study interfaces and device applications \cite{2018interface,2017natureenergy,2016materialstoday}. Van der Waals heterostructures are found to be an exotic platform that could realize fascinating phenomena \cite{Solis-Fernandez2017,2013MoS2-MoSe2,Cho2015,2019nc-hetero}. The emergence of 2D magnetic materials has facilitated to explore more functional heterostructures, e.g. multiferroicity \cite{2019nc-CGT} and tunable electronic structure \cite{2019GeS-FeCl2}.

\begin{figure*}[tbp]
\includegraphics[width=16.5cm]{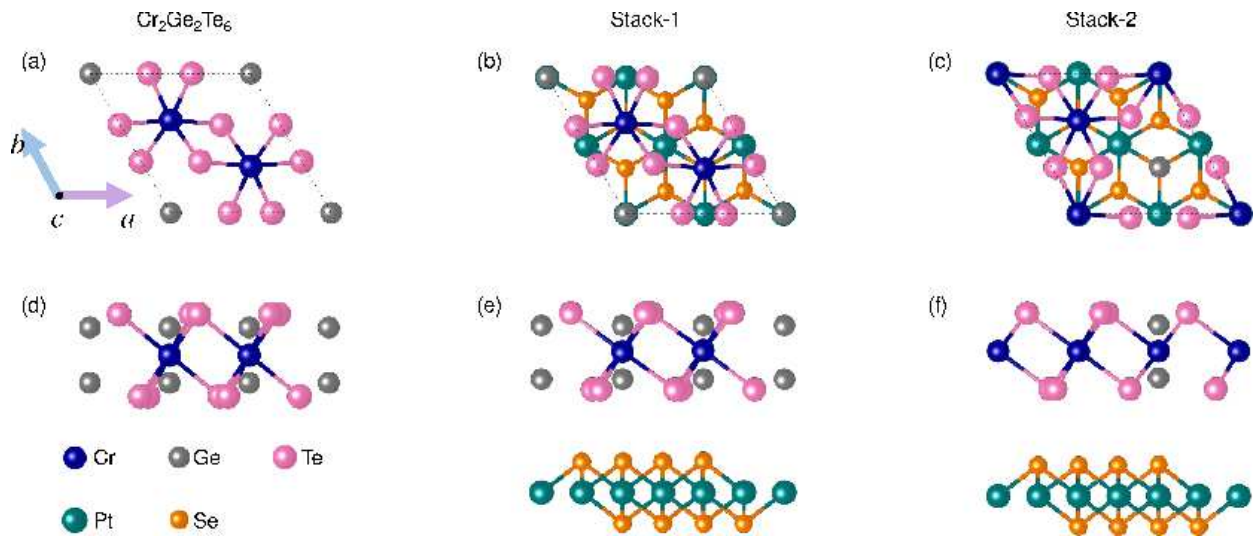}
\caption{ Crystal structures of two-dimensional Cr$_2$Ge$_2$Te$_6$ and Cr$_2$Ge$_2$Te$_6$/PtSe$_2$ heterostructures with Stack-1 and Stack-2, (a)-(c): top view; (d)-(f) side view, respectively.}
\label{F-1}
\end{figure*}

In WSe$_2$/CrI$_3$ heterostructures, the degenerate valleys in WSe$_2$ are split due to the proximity effect, where a large magnetic field of 13 T can be obtained by means of the ferromagnetic layer CrI$_3$ \cite{sciadv-CrI3-WSe2-2017}. The chiral edge state is manipulated by constructing WTe$_2$/bilayer CrI$_3$/WTe$_2$ heterostructures, in which CrI$_3$ was used to magnetize the topological insulator WTe$_2$ \cite{Chen2019edge}. In FeI$_2$/In$_2$Se$_3$ heterostructures, it is found that FeI$_2$ could undergo phase transition from ferromagnetic to antiferromagnetic (AFM) phase when the direction of ferroelectric polarization of In$_2$Se$_3$ is reversed \cite{Sun2019FeI}. Strong magnetic proximity effect via the s-d exchange coupling was proposed at the InAs/(Ga,Fe)Sb interface \cite{Takiguchi2019}. The effect of magnetic field on magnetic properties of metallic heterostructures Fe$_3$O$_4$/Pt was also studied \cite{2019Fe3O4-Pt}. The lattice mismatch is important for constructing heterostructures in experiments, and the epitaxial methods were often applied to produce van der Waals for highly lattice-mismatched systems \cite{Koma1992}. The heterostructures with large lattice mismatch about 10\% can be realized in the experiments. For example, the heterostructures GaSe/MoSe$_2$ with lattice mismatch of 13\% \cite{2016GaSe} and FeSe/Bi$_2$Te$_3$ with lattice mismatch of 19\% \cite{2017NPGlatticemismatch} were realized in recent experiments.

By constructing 2D van der Waals semiconductor heterostructures, is it possible to enhance the MAE and $T_C$ of ferromagnetic semiconductors? To answer this question, in this work, in terms of first principles calculations, we propose an efficient way to enhance the MAE and $T_C$ of Cr$_2$Ge$_2$Te$_6$, through constructing van der Waals heterostructures with nonmagnetic PtSe$_2$. Both Cr$_2$Ge$_2$Te$_6$ and PtSe$_2$ are 2D van der Waals semiconductors that have already been fabricated in experiments \cite{CGT-nature2017,2018ncPtSe2}. We consider two stack configurations and find the Curie temperatures $T_C$ of two stacks are larger than 600 K, while the MAE is increased by 70\%. We present a model and show that both the Dzyaloshinskii-Moriya interaction and the SIA contribute to the enhancement of the magnetic anisotropy. The increase of $T_C$ is partially owing to the decreased energy difference between 3d orbitals of Cr and 5p orbitals of Te. Our work will inspire further studies to enhance the magnetic properties of low-dimensional magnetic materials with the help of 2D van der Waals heterostructures.
Considering the great progress of 2D van der Waals stacking structures in recent experiments, such as the van der Waals stacking-dependent interlayer \cite{2019science-interlayer}, and general synthesis of 2D van der Waals heterostructure arrays \cite{2020duanxiangfeng}, our present proposal will be readily feasible for experiments.

\section{Calculation Methods}
We perform first principles calculations using Vienna $ab$ $initio$ simulation package (VASP) \cite{Kresse1996vasp}. Spin-polarized calculations are conducted with the Perdew-Burke-Ernzerhof exchange-correlation function and the generalized-gradient approximation. We adopted the DFT-D2 method of Grimme to describe the van der Waals interaction between Cr$_2$Ge$_2$Te$_6$ and PtSe$_2$ layers \cite{Grimme2006}. Spin-orbit coupling is taken into account in the calculations. 5 $\times$ 5 $\times$ 1 and 3 $\times$ 3 $\times$ 1 Mohnkhorst-Pack $k$-mesh sampling grids are used for the unit cell and supercell, respectively. The vacuum length is taken as 20 \AA, which is enough to isolate the present 2D system. The in-plane supercell is taken as $\sqrt{3}$ $\times$ 1 in a rectangular shape. The Hubbard $U$ for 3d electrons of Cr is chosen as 4 eV, which should be reasonable \cite{Dong2019}. We relaxed the structure of Cr$_2$Ge$_2$Te$_6$ and PtSe$_2$ to obtain the ground-state structure.
Based on the DFT results, the Curie temperature is calculated by using the Monte Carlo simulations \cite{Liu2016} based on the 2D Ising model, where a 60 $\times$ 60 supercell is adopted, and 10$^5$ steps are performed for every temperature to acquire the equilibrium.

\begin{table*}
\renewcommand\arraystretch{1.25}
\caption{The DFT results of lattice parameter and electronic and magnetic properties of considered Cr$_2$Ge$_2$Te$_6$ with various lattice constants, and Cr$_2$Ge$_2$Te$_6$/PtSe$_2$ heterostructures with Stack-1 and Stack-2 obtained by the GGA + SOC + $U$ calculations, where U = 4 eV.}
\begin{tabular}{lp{1.5cm}<{\centering}|p{2cm}<{\centering}p{2cm}<{\centering}p{2cm}<{\centering}|p{2cm}<{\centering}p{2cm}<{\centering}}
\hline
\hline
  & \multicolumn{4}{c|}{Cr$_2$Ge$_2$Te$_6$} & \multicolumn{2}{c}{Cr$_2$Ge$_2$Te$_6$/PtSe$_2$}\\
  \hline
  & Expt.  & Expt. lat. &  Relax lat. & Strain lat. &  Stack-1 & Stack-2 \\
\hline
lattice (\AA)         & 6.83  & 6.83      & 6.96     & 7.22     & 7.22      & 7.22 \\
band gap (eV)           & -     & 0.03      & 0.14     & 0.36     & 0.39      & 0.45 \\
$E_{FMz}$ (eV)          & -     & -44.5662  & -44.6458 & -44.4594 & -109.3624 & -109.2307 \\
$E_{FMx}$ (eV)          & -     & -44.5650  & -44.6445 & -44.4575 & -109.3601 & -109.2284 \\
$E_{AFM}$ (eV)          & -     & -44.4953  & -44.5399 & -44.3027 & -109.1711 & -109.0484  \\
S$_d$($\mu_B$)(Cr)      & -     & 3.59      & 3.52     & 3.66     & 3.59      & 3.68 \\
MAE (meV)               & -     & 1.19      & 1.33     & 1.89     & 2.32      & 2.23 \\
$T_C$ (K)               & 28    & 30        & 200      & 464      & 693       & 625 \\
\hline
\hline
\end{tabular}
\label{T-1}
\end{table*}

\section{Cr$_2$Ge$_2$Te$_6$/PtSe$_2$ heterostructures and enhancement of magnetism}

Bulk Cr$_2$Ge$_2$Te$_6$ has been synthesized in 1995 \cite{1995CGT} and successfully exfoliated into layers recently \cite{CGT-nature2017}. The space-group of 2D Cr$_2$Ge$_2$Te$_6$ is P$\overline{3}$1m (No.162). Monolayer PtSe$_2$ is considered as the most appropriate candidate to combine with 2D Cr$_2$Ge$_2$Te$_6$, which is a semiconductor with a lattice constant of 3.76 \AA\ that is well matched with the lattice of Cr$_2$Ge$_2$Te$_6$ and can be obtained by mechanical exfoliation methods \cite{2018ncPtSe2}. The space group of PtSe$_2$ is P$\overline{3}$m1 (No. 164), very similar to that of Cr$_2$Ge$_2$Te$_6$. The heterostructure is constructed in such a way that monolayer PtSe$_2$ with a size of 2 $\times$ 2 is attached to monolayer Cr$_2$Ge$_2$Te$_6$ with a size of 1 $\times$ 1. The lattice mismatch between Cr$_2$Ge$_2$Te$_6$ and PtSe$_2$ is 10\%, which could be executable in experiment for 2D van der Waals materials  heterostructures \cite{2019nature-vdw}. We propose two ways to make the heterostructures named as Stack-1 and Stack-2, in both configurations PtSe$_2$ is located under Cr$_2$Ge$_2$Te$_6$. In Stack-1, the center of Cr$_2$Ge$_2$Te$_6$ unit cell and the 2 $\times$ 2 PtSe$_2$ is aligned in $c$ direction, while Stack-2 can be treated as Stack-1 translating Cr$_2$Ge$_2$Te$_6$ layer with 1/3 unit cell along [1 2 0] direction. The space-group for Stack-1 and Stack-2 structures is P3 (No.143). The crystal structure of monolayer Cr$_2$Ge$_2$Te$_6$ and the two stacks are shown in Fig. \ref{F-1}. It is noted that the lattice constant becomes 7.22 \AA\ for the heterostructure Cr$_2$Ge$_2$Te$_6$/PtSe$_2$, which is equivalent to applying 3.7\% tensile strain to monolayer Cr$_2$Ge$_2$Te$_6$.

The properties of Cr$_2$Ge$_2$Te$_6$ are sensitive to its lattice constant \cite{2019nc-CGT}. For 2D Cr$_2$Ge$_2$Te$_6$, the lattice constant is 6.83 \AA\ in the experiment \cite{CGT-nature2017}, and becomes 6.96 \AA\ in the optimized structure. The experimental and the calculated results with experimental and relaxed lattice constants are given in Table \ref{T-1}. The experimental data of T$_C$ and lattice constant for 2D Cr$_2$Ge$_2$Te$_6$ shown in Table \ref{T-1} are taken from Ref. \cite{CGT-nature2017}. For the hexagonal lattice, we used the following three configurations for AFM ordering: the AFM $N\acute{e}el$, AFM stripe, and AFM zigzag states \cite{Dong2019}. The AFM state used for discussions in Table \ref{T-1} is the AFM zigzag state, which is the lowest energy AFM state according to our calculations. In addition, for the bulk Cr$_2$Ge$_2$Te$_6$, there are some experimental reports, for example, the MAE is 0.25 meV per unit cell of Cr$_6$Ge$_6$Te$_{18}$ \cite{Zeisner2019}, T$_C$ is about 66 K \cite{Zeisner2019} and 62.5 K \cite{2017prbliu}, and the spin moment is 3.41 $\mu_B$ \cite{2017prbliu}. 

The MAE is defined by the energy difference of the in-plane and the out-of-plane ferromagnetic state and can be expressed as $MAE = E_{FMx} - E_{FMz}$, where $E_{FMx}$ and $E_{FMz}$ denote the energy of the ferromagnetic state with in-plane and out-of-plane direction of the unit cell, respectively. Our results suggest that these materials possess a ferromagnetic ground state with out-of-plane magnetization.

\begin{figure*}[tbp]
	\includegraphics[width=16.5cm]{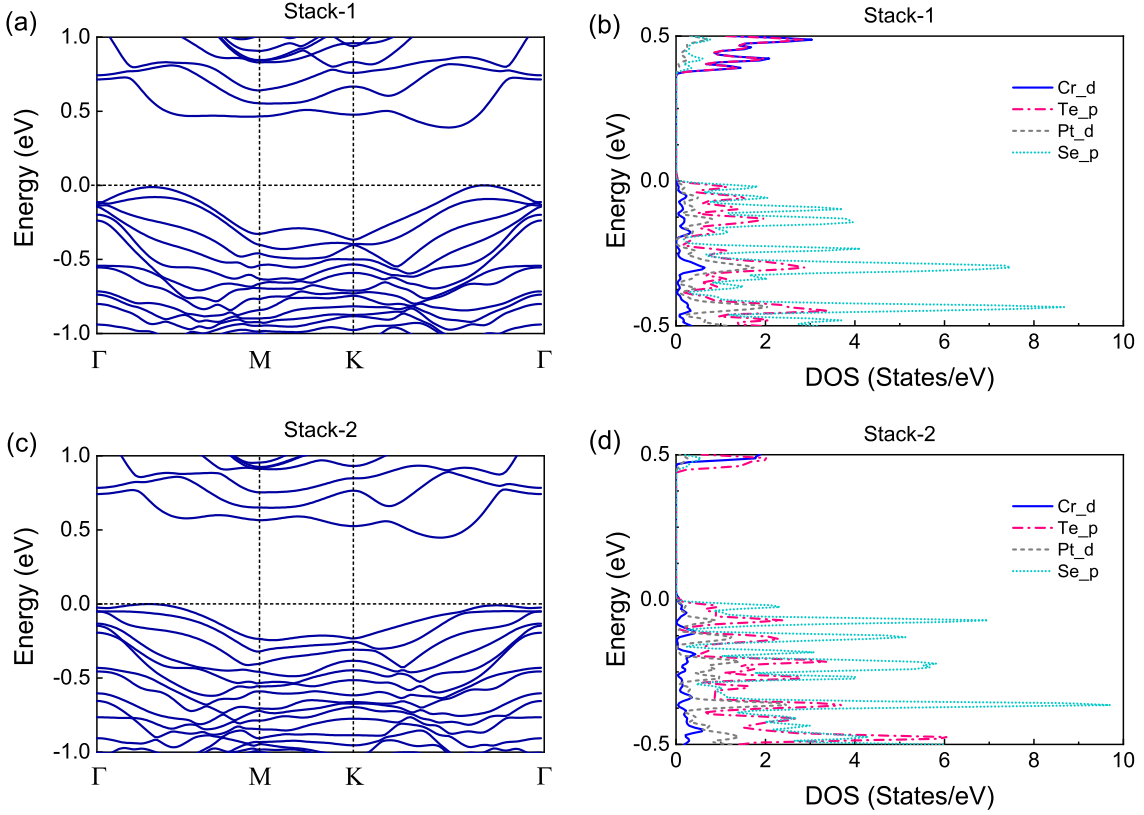}
	\caption{Electronic band structures and partial DOS of two-dimensional Cr$_2$Ge$_2$Te$_6$/PtSe$_2$ heterostructures with (a), (b) Stack-1 and (c), (d) Stack-2, respectively, obtained by the GGA + SOC + $U$ calculations with U = 4 eV.}
	\label{F-2}
\end{figure*}

Our calculations show that Cr$_2$Ge$_2$Te$_6$/PtSe$_2$ heterostructure Stack-1 and Stack-2 are semiconductors with finite band gaps of 0.39 and 0.45 eV, respectively. The electronic band structures and atom-projected partial density of states of Stack-1 and Stack-2 are presented in Fig. \ref{F-2}.

\begin{figure}[tbp]
\includegraphics[width=8.5cm]{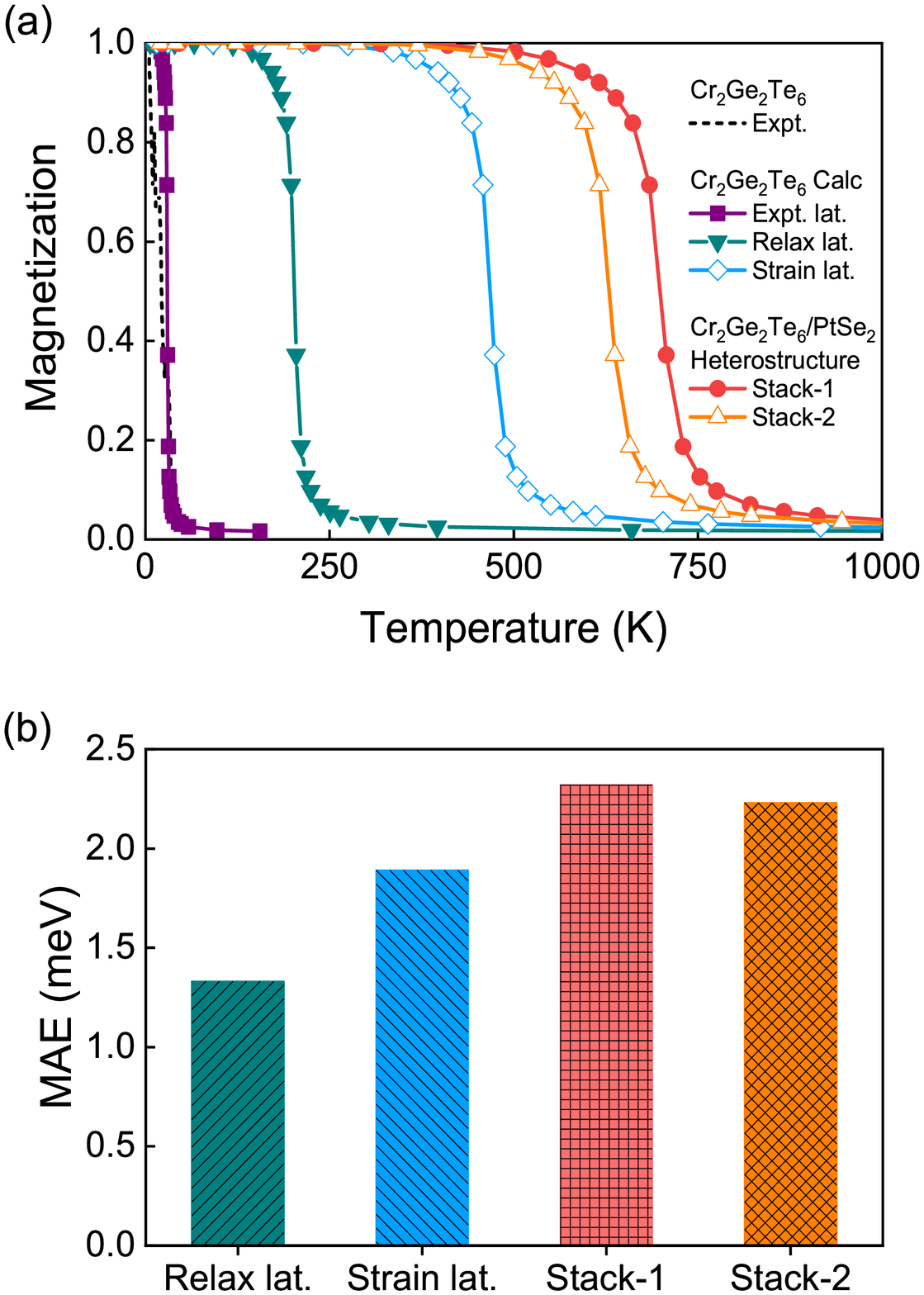}
\caption{For two-dimensional Cr$_2$Ge$_2$Te$_6$, Cr$_2$Ge$_2$Te$_6$ with strain, and Cr$_2$Ge$_2$Te$_6$/PtSe$_2$ heterostructures with Stack-1 and Stack-2, (a) the normalized magnetization as a function of temperature, in which Cr$_2$Ge$_2$Te$_6$ with experimental lattice was taken from our previous work \cite{Dong2019} and (b) the magnetic anisotropic energy (MAE). The experimental result of Cr$_2$Ge$_2$Te$_6$ is taken from Ref. \cite{CGT-nature2017}. The calculated results are obtained by the GGA + SOC + $U$ calculations and the Monte Carlo simulations, where U = 4 eV.}
\label{F-3}
\end{figure}

The magnetization as a function of temperature obtained by Monte Carlo simulations based on 2D Ising model is depicted in Fig. \ref{F-3}(a). One may note that our calculated result for $T_C$ of monolayer Cr$_2$Ge$_2$Te$_6$ agrees well with the experimental result, revealing that our calculating method is reliable.  It is shown that the Curie temperature $T_C$ of Cr$_2$Ge$_2$Te$_6$ with 3.7\% tensile strain (the lattice constant 7.22 \AA) is increased beyond room temperature, and $T_C$ of the heterostructures with Stack-1 and Stack-2 is increased above 600 K. In addition to the increase of Curie temperature, the MAE is also increased by about 40\% in the Cr$_2$Ge$_2$Te$_6$ with 3.7\% tensile strain, and enhanced by about 70\% in the heterostructures with Stack-1 and Stack-2 as shown in Fig. \ref{F-3}(b). It is indicated that the MAE of Stack-1 and Stack-2 are both larger than that of Cr$_2$Ge$_2$Te$_6$ monolayer, showing that the nonmagnetic layer PtSe$_2$ has a remarkable effect on the enhancement of magnetism in 2D Cr$_2$Ge$_2$Te$_6$, which will be discussed in next section.

\begin{figure}[tbp]
\includegraphics[width=8.5cm]{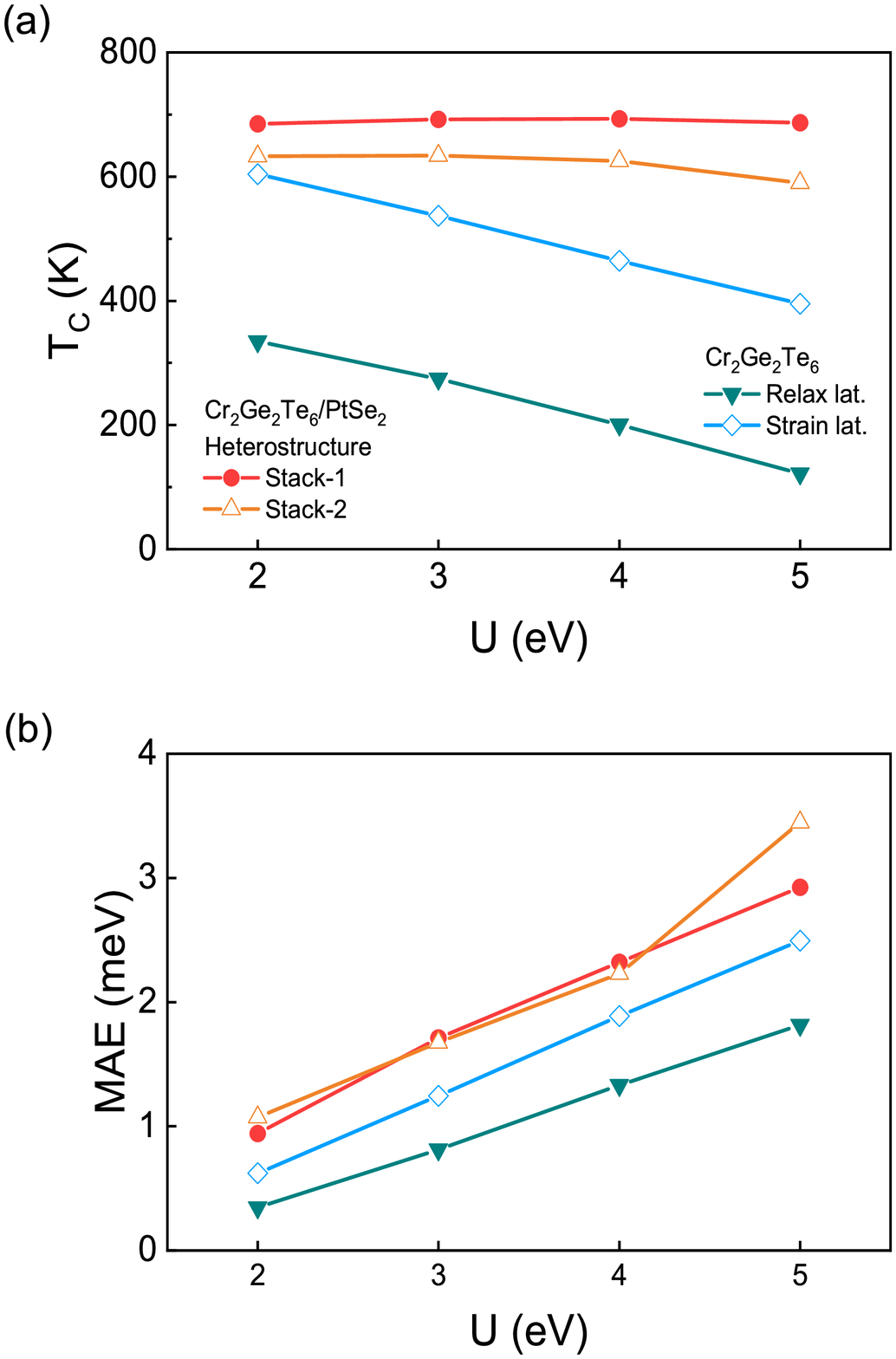}
\caption{The Hubbard parameter U dependence of (a) Curie temperature $T_C$ and (b) magnetic anisotropic energy (MAE) for Cr$_2$Ge$_2$Te$_6$-relaxed, Cr$_2$Ge$_2$Te$_6$-strain (3.7\%), Cr$_2$Ge$_2$Te$_6$/PtSe$_2$ heterostructures with Stack-1 and Stack-2, obtained by the GGA + SOC + $U$ calculations.}
\label{F-4}
\end{figure}

To study the effect of Hubbard parameter U on the calculated $T_C$ and MAE, we perform DFT calculations for different values of U, as shown in Fig. \ref{F-4}. One may see that with increasing U, $T_C$ of Stack-1 and Stack-2 remain almost unaltered, while $T_C$ of monolayer Cr$_2$Ge$_2$Te$_6$ is decreasing with the increase of U. However, with increasing U the MAE is found to increase for monolayer Cr$_2$Ge$_2$Te$_6$ and heterostructure Stack-1 and Stack-2, revealing that the enhancements of $T_C$ and MAE have different mechanisms, and electron interactions play important roles in the enhancement of MAE. In addition, the U-dependent band gap is also studied by DFT calculations. Our results show that the band gap slightly increases with increasing U in the range of 4 - 5 eV, as this is typical for 3d electrons.

\section{Theoretical Analysis}
To describe the magnetic anisotropy in monolayer Cr$_2$Ge$_2$Te$_6$ and the heterostructure Cr$_2$Ge$_2$Te$_6$/PtSe$_2$, we write down a general spin Hamiltonian

\begin{equation}
\begin{split}
\hat{H}_{spin} = \sum_{<i,j>}J_{xx}S_{ix}S_{jx} + \sum_{<i,j>}J_{xy}S_{ix}S_{jy} + \sum_{<i,j>}J_{xz}S_{ix}S_{jz} \\
+ \sum_{<i,j>}J_{yx}S_{iy}S_{jx} + \sum_{<i,j>}J_{yy}S_{iy}S_{jy} + \sum_{<i,j>}J_{yz}S_{iy}S_{jz}\\
+ \sum_{<i,j>}J_{zx}S_{iz}S_{jx} + \sum_{<i,j>}J_{zy}S_{iz}S_{jy} + \sum_{<i,j>}J_{zz}S_{iz}S_{jz} \\
+ \sum_{i,\alpha,\beta} A_{\alpha \beta}S_{i\alpha}S_{i\beta}.
\end{split}
\end{equation}

In Eq. (1), $i$ and $j$ denote the magnetic atom sites, and the summation on $<i,j>$ only includes the nearest neighbors. $\alpha$ and $\beta$ denote three directions of $x$ $y$ and $z$ axis. The former nine terms in Eq. (1) are the nearest neighbor exchange interactions between different spin components and the last term represents SIA energy. $J_{\alpha\beta}$ are the matrix elements of exchange couplings between the nearest spins, and the off-diagonal elements lead to the Kitaev interactions \cite{npjkitaev2018,2018-K-naturephy} and the Dzyaloshinskii-Moriya (DM) \cite{2019DM-nature-mat,2014DM-prb,2019DM-nanotech} interactions. 
Matrix elements of exchange coupling matrix J of Cr$_2$Ge$_2$Te$_6$ and Cr$_2$Ge$_2$Te$_6$/PtSe$_2$ could be calculated by the GGA + SOC + $U$ method and the four-state method \cite{2013dalton}, where we take, as usual, U = 4 eV.
\begin{table}
\renewcommand\arraystretch{1.25}
\caption{The diagonal elements of exchange coupling matrix J (in units of meV) for monolayer Cr$_2$Ge$_2$Te$_6$, 2D Cr$_2$Ge$_2$Te$_6$ with 3.7\% tensile strain and Cr$_2$Ge$_2$Te$_6$/PtSe$_2$ heterostructures calculated by the GGA + SOC + $U$ and the four-state method \cite{2013dalton}. Here U = 4 eV.}
\begin{tabular}{lp{2cm}<{\centering}p{2cm}<{\centering}p{2cm}<{\centering}}
\hline
\hline
  & J$_{xx}$ & J$_{yy}$  & J$_{zz}$  \\
\hline
Cr$_2$Ge$_2$Te$_6$-relax  & -8.068  & -7.467  & -8.316  \\
Cr$_2$Ge$_2$Te$_6$-strain & -11.314 & -10.591 & -11.620 \\
Stack-1 & -12.557 & -11.813 & -12.896 \\
Stack-2 & -12.290 & -11.508 & -12.623\\
\hline
\hline
\end{tabular}
\label{T-2}
\end{table}	

To calculate parameter J$_{\alpha\beta}$, for instance, J$_{xy}$, the following four classical spin configurations are used \cite{2013dalton}:
 $S_1$ = (1, 0, 0), $S_2$ = (0, 1, 0); $S_1$ = (1, 0, 0), $S_2$ = (0, -1, 0); $S_1$ = (-1, 0, 0), $S_2$ = (0, 1, 0); $S_1$ = (-1, 0, 0), $S_2$ = (0, -1, 0). Keep the same for the rest spins, where $S$ = (0, 0, 1) is used in our calculation. Eq. (1) with these four spin configurations can be written as
\begin{flalign}
\begin{split}
&E_1 = J_{xy}S_{1x}S_{2y} + E_0 = J_{xy}|S|^2 + E_0,    \\
&E_2 = -J_{xy}S_{1x}S_{2y} + E_0 = -J_{xy}|S|^2 + E_0,  \\
&E_3 = -J_{xy}S_{1x}S_{2y} + E_0 = -J_{xy}|S|^2 + E_0,  \\
&E_4 = J_{xy}S_{1x}S_{2y} + E_0 = J_{xy}|S|^2 + E_0.
\end{split}
\end{flalign}
One has $J_{xy}=((E_1+E_4)-(E_2+E_3 ))/4$.
Other elements of J can be obtained in a similar way.
 The diagonal elements of J are given in Table \ref{T-2}. The negative values mean the ferromagnetic coupling. It is clear that the ground state is ferromagnetic for these four structures. The off-diagonal elements of J and their differences of monolayer Cr$_2$Ge$_2$Te$_6$ and Cr$_2$Ge$_2$Te$_6$/PtSe$_2$ heterostructures are given in Table \ref{T-3}. 
 
 The magnitude of DM interaction can be estimated as 
 \begin{equation}
 D_{z} = \frac{|J_{yx} - J_{xy}|}{2}.
 \end{equation}

 \begin{table*}
\renewcommand\arraystretch{1.25}
\caption{The off-diagonal elements of exchange coupling matrix J (in units of meV) for monolayer Cr$_2$Ge$_2$Te$_6$ with relaxation and strain, and Cr$_2$Ge$_2$Te$_6$/PtSe$_2$ heterostructures with Stack-1 and Stack-2 calculated by the GGA + SOC + $U$ method and the four-state method \cite{2013dalton}. Here U = 4 eV.}
\begin{tabular}{lp{1.5cm}<{\centering}p{1.5cm}<{\centering}p{1.5cm}<{\centering}p{1.5cm}<{\centering}p{1.5cm}<{\centering}p{1.5cm}<{\centering}ccc}
\hline
\hline
  & J$_{xy}$ & J$_{yx}$  & J$_{xz}$  & J$_{zx}$ & J$_{yz}$ & J$_{zy}$  & $|$J$_{xy}$ - J$_{yx}$ $|$ & $|$J$_{xz}$ - J$_{zx}$ $|$ &  $|$J$_{yz}$ - J$_{zy}$ $|$ \\
\hline
Cr$_2$Ge$_2$Te$_6$-relax   & -0.493 & -0.493 & 0.069  & 0.069  & -0.111 & -0.111 & 0.000 & 0.000 & 0.000   \\
Cr$_2$Ge$_2$Te$_6$-strain  & -0.593 & -0.593 & -0.061 & -0.061 & 0.102  & 0.102  & 0.000 & 0.000 & 0.000  \\
Stack-1 & -0.750 & -0.539  & -0.053 & 0.228  & -0.188 & -0.072 & 0.211  & 0.281  & 0.116   \\
Stack-2 & 0.669  & 0.628   & -0.153 & 0.334  & 0.293  & 0.014  & 0.041  & 0.487  & 0.279 \\
\hline
\hline
\end{tabular}
\label{T-3}
\end{table*}

 We find that for monolayer Cr$_2$Ge$_2$Te$_6$, J$_{xy}$ = J$_{yx}$, J$_{xz}$ = J$_{zx}$, J$_{zy}$ = J$_{yz}$, and thus there is no DM interaction. This is expected, because the inversion symmetry is kept in the monolayer. For the Cr$_2$Ge$_2$Te$_6$/PtSe$_2$ heterostructures, the difference of the off-diagonal elements are finite values, which implies the presence of DM interaction. 

\begin{figure}[tbp]
\includegraphics[width=8.5cm]{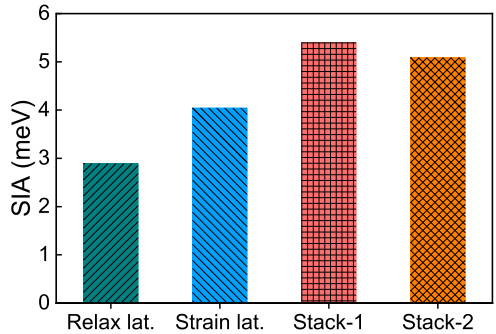}
\caption{The single-ion anisotropy (SIA) energy for Cr$_2$Ge$_2$Te$_6$-relaxed, Cr$_2$Ge$_2$Te$_6$-strain (3.7\%) Cr$_2$Ge$_2$Te$_6$/PtSe$_2$ heterostructures with Stack-1 and Stack-2 obtained by the GGA + SOC + $U$ calculations, with U = 4 eV.}
\label{F-5}
\end{figure}

The other contribution to the MAE is the SIA energy, i.e. the last term in Eq. (1). For monolayer Cr$_2$Ge$_2$Te$_6$ with relaxation and strain, and Cr$_2$Ge$_2$Te$_6$/PtSe$_2$ heterostructure with Stack-1 and Stack-2, the SIA energy is calculated by the GGA + SOC + $U$ method, where we assume again U = 4 eV. As shown in Fig. \ref{F-5}, the SIA energy increases from monolayer Cr$_2$Ge$_2$Te$_6$ to Stack-1 and Stack-2. Comparing the SIA in Fig. \ref{F-5} with the MAE in Fig. \ref{F-3}(b), one may note that the enhancement of the SIA and MAE is very similar.

The above DM interaction and the SIA originate from the spin-orbit coupling, which in turn contribute to the MAE. For recent magnetic semiconductor monolayers, various microscopic mechanisms of MAE are also discussed \cite{2019MRS,2019prlRuCl3,2019prmaterials}.

We also calculate the second nearest neighbor exchange coupling J$_2$ and the nearest neighbor exchange coupling J$_1$ by comparing different magnetic configurations \cite{2019guoguangyu,2018nanolett}. For 2D Cr$_2$Ge$_2$Te$_6$ with relaxation, we obtain J$_1$ = -8.38 meV, J$_2$ = 0.43 meV; for 2D Cr$_2$Ge$_2$Te$_6$ with strain, J$_1$ = -11.64 meV, J$_2$ = -0.06 meV; for Cr$_2$Ge$_2$Te$_6$/PtSe$_2$ Stack-1, J$_1$ = -13.01 meV, J$_2$ = -0.42 meV; for Cr$_2$Ge$_2$Te$_6$/PtSe$_2$ Stack-2, J$_1$ = -12.75 meV, J$_2$ = -0.36 meV. Our results show that J$_2$ was about 20 to 30 times smaller than J$_1$ in magnitude in present cases. This is in contrast with the results in 2D CrI$_3$, where the calculated J$_2$ is about 5 times smaller than J$_1$ in magnitude \cite{2019guoguangyu}. It is noted that $T_C$ is determined by the leading term J$_1$, and barely affected by small values of J$_2$. Thus, the second nearest neighbor exchange coupling can be neglected in our case. Here we should mention that for the simplicity we applied an isotropic Heisenberg model to get J$_1$ values by comparing the energies between different magnetic configurations. Thus, the values of J$_1$ are approximately equal to the average of J$_{xx}$, J$_{yy}$ and J$_{zz}$.

In fact, the above results of J$_2$/J$_1$ are consistent with the magnetic anisotropy in the experiments of 2D CrI$_3$ and Cr$_2$Ge$_2$Te$_6$ \cite{CGT-nature2017,CrI3-2017}. For the 2D CrI$_3$, J$_2$/J$_1$ $\sim$ 0.2 was obtained in the calculation \cite{2019guoguangyu}, and thus the DM interaction can be expected, which is consistent with the large magnetic anisotropy in 2D CrI$_3$ in the experiment. In contrast, for the 2D Cr$_2$Ge$_2$Te$_6$, a very small J$_2$/J$_1$ $\sim$ 0.05 was obtained in our calculation, which is also consistent with the small magnetic anisotropy in the experiment of 2D Cr$_2$Ge$_2$Te$_6$. Because the MAE in the experiment of 2D Cr$_2$Ge$_2$Te$_6$ is weak, we propose a scheme here to enhance MAE by constructing the heterostructure Cr$_2$Ge$_2$Te$_6$/PtSe$_2$, where the inversion symmetry is broken, and the DM interaction is expected.

\begin{table*}
\renewcommand\arraystretch{1.25}
\caption{The DFT results of mixing-matrix $|V_{pd}|$ and energy difference $|E_p$ - $E_d|$ between 5$p$ orbitals of Te and 3$d$ orbitals of Cr for Cr$_2$Ge$_2$Te$_6$ relaxed and strained (in units of eV).}
\begin{tabular}{lp{3cm}<{\centering}p{3cm}<{\centering}p{3cm}<{\centering}p{3cm}<{\centering}p{3cm}<{\centering}}
\hline
\hline
  & \multicolumn{5}{c}{Cr$_2$Ge$_2$Te$_6$-relax} \\
\cline{2-6}
  & $p_y$ - $d_{x^2-y^2}$ & $p_y$ - $d_{z^2}$ & $p_x$ - $d_{xy}$ & $p_y$ - $d_{yz}$ & $p_z$ - $d_{yz}$  \\
\hline
$|V_{pd}|$          & 0.7099 & 0.3583 & 0.3287 & 0.2879 & 0.2793  \\
$|E_p$ - $E_d|$     & 0.6659 & 0.3525 & 1.1036 & 0.9681 & 1.4864  \\
\hline
  & \multicolumn{5}{c}{Cr$_2$Ge$_2$Te$_6$-strain} \\
\cline{2-6}
  & $p_x$ - $d_{xy}$ & $p_y$ - $d_{xy}$ & $p_y$ - $d_{xz}$ & $p_z$ - $d_{xy}$ & $p_x$ - $d_{yz}$  \\
\hline
$|V_{pd}|$          & 0.4061 & 0.4021 & 0.2960 & 0.2829 & 0.2561  \\
$|E_p$ - $E_d|$     & 0.2943 & 0.3379 & 0.4141 & 0.3030 & 0.2888  \\
\hline
\hline
\end{tabular}
\label{T-5}
\end{table*}		

\section{Discussion}

To describe the magnetic state of 2D Cr$_2$Ge$_2$Te$_6$ and the heterostructures Cr$_2$Ge$_2$Te$_6$/PtSe$_2$, the Heisenberg model is used, as defined in Eq. (1). To estimate some magnetic properties, such as Curie temperature $T_C$ and exchange parameters, the approximations with classical spins are used. For example, to calculate $T_C$, we carry out the Monte Carlo simulations based on the Ising model. To calculate the exchange parameters J$_{xx}$, J$_{xy}$, etc. in Eq. (1), the four-state method with four classical spin configurations are used \cite{2013dalton}.

As shown in Fig. \ref{F-4}(a) and (b), the Hubbard on-site Coulomb correlation U dependence of $T_C$ and MAE are calculated. For U = 1 eV, our DFT results find that 2D Cr$_2$Ge$_2$Te$_6$ has a ferromagnetic ground state with in-plane magnetization (negative MAE, not shown in Fig. \ref{F-4}), in contrast to the experimental observation that the out-of-plane magnetization was observed. Thus, this result suggests that the small U = 1 eV was not reasonable for 2D Cr$_2$Ge$_2$Te$_6$. Instead, our DFT calculations suggest that a large parameter U = 4 eV is reasonable for 2D Cr$_2$Ge$_2$Te$_6$, where both $T_C$ and MAE are consistent with the experimental results. In Fig. \ref{F-4}(a) and (b), $T_C$ and MAE with U ranging from 2 - 5 eV are presented. It is shown that our conclusions, i.e., the great enhancement of Curie temperature and magnetic anisotropy in 2D van der Waals magnetic semiconductor heterostructures, do not change with Hubbard U in the range of 2 - 5 eV.

The great increase of $T_C$ in Cr$_2$Ge$_2$Te$_6$ by applying tensile strain through heterostructures could be understood based on the superexchange interaction \cite{Goodenough1955,Kanamori1960,Anderson1959}. To deal with the superexchange interaction between the nearest neighbor Cr atoms we adopted the simple four-electron model \cite{2017ferromag}. In this model, four electrons d$_1$, p, p$^\prime$, d$_2$ are electrons of Cr$_1$, Te (contains two p electrons), and Cr$_2$ atoms, respectively. The indirect magnetic interaction between
Cr$_1$ and Cr$_2$ can be approximately written as $J_{indirect} = [1/E(\uparrow \downarrow)^2 -1/E(\uparrow \uparrow)^2]b^2J_{pd}$. The superexchange process could be divided into two intermediate processes. One intermediate process is the transfer of the p electron of Te to Cr$_1$ site, $E(\uparrow \downarrow)$ is the energy needed if spins of p and d$_1$ electrons form a singlet state, $E(\uparrow \uparrow)$ is the energy needed if spins of p and d$_1$ electrons form a triplet state, and b is the integral of this intermediate process. The other intermediate process is the direct exchange coupling between the remaining p$^\prime$ electron of Te and the d$_2$ electron of Cr$_2$, where the antiferromagnetic p-d exchange coupling J$_{pd}$ can be approximately expressed as $J_{pd} = 2|V_{pd}|^2\frac{U}{(E_p - E_d)(E_d - E_p + U)}$ following the s-d exchange model with Schriffer-Wolff transformation \cite{Schrieffer1966}. For large U $\gg$ $|E_p - E_d|$, J$_{pd}$ could be simplified to $J_{pd} \approx 2|V_{pd} |^2\frac{1}{|E_p - E_d|}$. V$_{pd}$ is the mixing matrix element between 5p orbitals of Te and 3d orbitals of Cr, and ($E_p - E_d$) is the energy difference between 5p orbitals of Te and 3d orbitals of Cr. It is noted that both mixing term V$_{pb}$ and energy difference  $|E_p - E_d|$ are materials dependent.

By DFT calculations, we can obtain these parameters for 2D Cr$_2$Ge$_2$Te$_6$ relaxed and with a strain. The results of $|V_{pd}|$ and $|E_p - E_d|$ are listed in Table \ref{T-5}. Comparing the results of Cr$_2$Ge$_2$Te$_6$ with relaxation and Cr$_2$Ge$_2$Te$_6$ with strain (3.7\% tensile strain), we find that the increased antiferromagnetic coupling J$_{pd}$ by strain is mainly from the decreased energy difference $|E_p - E_d|$. The decreased $|E_p - E_d|$ gives rise to the enhanced J$_{pd}$, J$_{indirect}$ and $T_C$. There is no essential difference for the Cr-Te-Cr bond angles before and after applying the strain, where Cr-Te-Cr bond angle is 89.8 degree for Cr$_2$Ge$_2$Te$_6$-relax, and 92.3 degree for Cr$_2$Ge$_2$Te$_6$-strain. It is consistent with the mixing term $|V_{pd}|$ as shown in Table \ref{T-5}, where there is no essential difference for $|V_{pd}|$ before and after applying the strain.

As shown in Fig. \ref{F-4}(a), it is clear that the increased $T_C$ between the Cr$_2$Ge$_2$Te$_6$ with relax and the Cr$_2$Ge$_2$Te$_6$ with strain are nearly U independent, which is consistent with the nearly U independent J$_{pd}$. For the enhanced $T_C$ of heterostructures, as shown in Fig. \ref{F-4} (a), an important contribution comes from the strain effect in the heterostructures. For the $T_C$ in heterostructures Cr$_2$Ge$_2$Te$_6$/PtSe$_2$, other contribution rather than strain also appears with U $>$ 2 eV as shown in Fig. \ref{F-4}(a), while the reason of such contribution needs further studies.

The MAE increases with increasing U, as shown in Fig. \ref{F-4}(b). This behavior could be understood by the multiorbital electron correlations. The multiorbital Coulomb interactions between electrons with the orbitals \textbf{m} and -\textbf{m} and spin $\sigma$ is expressed as \cite{Nguyen2018,You2019}:
 $H_{m\sigma} = (U^\prime - J_H)n_{m\sigma}n_{m\overline{\sigma}} \approx (U^\prime - J_H)(\overline{n}^2 - \frac{1}{4}\lambda_{SO}^2m^2\sigma^2(\delta n)^2 )$,
 where U is the on-site Coulomb interactions within the same orbitals, U$^\prime$ is the on-site Coulomb interactions between different orbitals, and J$_H$ demotes the Hund coupling. In the atomic limit, these parameters satisfy the relation U = U$^\prime$ + 2J$_H$ \cite{Maekawa2004}. $\lambda_{SO}$ is the SOC parameter, giving rise to the band splitting due to spin-orbit coupling. To compensate the increased energy due to the multiorbital Coulomb interactions, the SOC will increase with increasing U. Such enhancement of SOC due to the electron correlations has been discussed in the large topological band gap \cite{You2019} and the enhancement of spin Hall effect \cite{gulaoshi2010,guoguangyu2009}. Because the MAE is originated from the SOC, MAE will also increase with increasing U.

When growing two materials together, two kinds of junctions can be considered. One is the incoherent junction, where each layer of the junction keeps their original lattice constants \cite{Koma1992}, and there is little strain at the junction. Another is the coherent junction, where the lattice constant of each layer is forced to be the same, and the strain is expected at the junction. In this work, the coherent junction is considered, and the strain at the junction is found to play an important role in enhancing the magnetic anisotropy and Curie temperature of heterostructures. Our results suggest that it is better to apply the coherent junctions to manipulate the magnetic properties of heterostructures.

\section {Summary}

In this paper, we have proposed that the magnetic anisotropic energy (MAE) of 2D ferromagnetic semiconductor can be strongly enhanced in heterostructures with a nonmagnetic semiconductor monolayer with large spin-orbit coupling. Based on the density functional theory calculations, we have demonstrated this idea in the bilayer heterostructure Cr$_2$Ge$_2$Te$_6$/PtSe$_2$. The results show that the MAE of Cr$_2$Ge$_2$Te$_6$ is enhanced by 70\%, and its Curie temperature is increased to larger than 600 K far beyond room temperature. By means of a model Hamiltonian analysis, it is shown that both the Dzyaloshinskii-Moriya interaction and the single-ion anisotropy contribute to the enhancement of the MAE. Based on the superexchange picture, the great enhancement of $T_C$ is partially attributed to the decreased energy difference between 3d orbitals of Cr and 5p orbitals of Te. Our present study gives a deeper understanding on the mechanism of enhancement of magnetism in 2D van der Waals magnetic/nonmagnetic heterostructures and will spur further experimental studies of the 2D magnetic semiconductor systems.

\section* {Acknowledgments}
The authors acknowledge Q. B. Yan, Z. G. Zhu, and Z. C. Wang for many valuable discussions. This paper is supported in part by the National Key R\&D Program of China (Grant No. 2018YFA0305800), the Strategic Priority Research Program of the Chinese Academy of Sciences (Grant No. XDB28000000), the National Natural Science Foundation of China (Grant
No.11834014), and Beijing Municipal Science and Technology Commission (Grant No. Z190011). B.G. is also supported by the National Natural Science Foundation of China (Grant No.Y81Z01A1A9), the Chinese Academy of Sciences (Grant No. Y929013EA2), the University of Chinese Academy of Sciences (Grant No. 110200M208), the Strategic Priority Research Program of Chinese Academy of Sciences (Grant No.XDB33000000),and the Beijing Natural Science Foundation (Grant No. Z190011).

%

\end{document}